\pdfoutput=1

\documentclass[11pt,twoside,a4paper,cmspaper,final,collab]{cms-tdr}
\def\svnVersion{104165:104171}\def\cmsCernNo{2012-033}\def\cmsCernDate{\today}\def\cmsMessage{Submitted to the Journal of High Energy Physics}

\begin{document}\cmsNoteHeader{HIG-11-026}

\hyphenation{had-ron-i-za-tion}
\hyphenation{cal-or-i-me-ter}
\hyphenation{de-vices}

\RCS$Revision: 104146 $
\RCS$HeadURL: svn+ssh://alverson@svn.cern.ch/reps/tdr2/papers/HIG-11-026/trunk/HIG-11-026.tex $
\RCS$Id: HIG-11-026.tex 104146 2012-02-14 22:53:25Z avartak $
\providecommand{\mH}{\ensuremath{m_{\PH}\xspace}}
\providecommand{\MT}{\ensuremath{M_{\mathrm{T}}\xspace}}
\cmsNoteHeader{HIG-11-026} % This is over-written in the CMS environment: useful as preprint no. for export versions
\title{Search for the standard model Higgs boson in the \texorpdfstring{$\PH \rightarrow \cPZ\cPZ \rightarrow 2\ell 2\nu$}{H to ZZ to 2l 2nu} channel in \Pp\Pp\ collisions at $\sqrt{s}$ = 7 TeV}

\date{\today}

\abstract{
A search for the standard model Higgs boson in the
$\PH \rightarrow \cPZ\cPZ \rightarrow 2\ell 2\cPgn$ decay channel, where $\ell = \Pe$ or $\Pgm$,
in \Pp\Pp\ collisions at a center-of-mass energy of 7\TeV is presented.
The data were collected at the  LHC, with the CMS detector,
and correspond to an integrated luminosity of 4.6\fbinv.
No significant excess is observed above the background expectation,
and upper limits are set on the Higgs boson production cross section.
The presence of the standard model Higgs boson with a mass in the 270--440\GeV range is excluded at 95\% confidence level.
}

\hypersetup{%
pdfauthor={CMS Collaboration},%
pdftitle={Search for the standard model Higgs boson in the H to ZZ to 2l 2nu channel in pp collisions at sqrt(s) = 7 TeV},%
pdfsubject={CMS},%
pdfkeywords={CMS, physics, higgs}}

\maketitle %maketitle comes after all the front information has been supplied

\section{Introduction}
\label{sec:introduction}

The standard model (SM) of particle physics
~\cite{Glashow:1961tr, Weinberg:1967tq, SalamNobel}
accommodates essentially all relevant experimental data.
One of the remaining questions is the origin of
mass for fundamental particles. Within the SM, vector boson masses arise from the
spontaneous breaking of electroweak symmetry~\cite{Englert:1964et, Higgs:1964ia, Higgs:1964pj, Guralnik:1964eu, Higgs:1966ev, Kibble:1967sv}.
The existence of the associated field quantum, the Higgs boson,
has yet to be experimentally established. The discovery or exclusion
of the SM Higgs boson is one of the main goals of the physics programme at the CERN Large
Hadron Collider (LHC).

To date, experimental searches for the SM Higgs boson have
yielded null results. Limits at 95\% confidence level (CL) on its mass ($\mH$)
have been placed by experiments at the Large Electron-Positron Collider (LEP),
$\mH > 114.4\GeV$~\cite{LEPHIGGS}, the Tevatron, $\mH \notin$ (162--166)\GeV~\cite{TEVHIGGS_2010},
and ATLAS, $\mH \notin$ (145--206), (214--224), and (340--450)\GeV~\cite{Aad:2011uq,ATLAS:2011af,ATLAS:2011aa}.
The primary production mechanism for the Higgs boson at the LHC is through gluon fusion
~\cite{Djouadi:1991tka,Dawson:1990zj,Spira:1995rr,Harlander:2002wh,Anastasiou:2002yz,Ravindran:2003um,Aglietti:2004nj,Degrassi:2004mx,Actis:2008ug, Anastasiou:2008tj,deFlorian:2009hc,Baglio:2010ae}
with a small contribution from vector boson fusion (VBF)~\cite{ Ciccolini:2007jr,Ciccolini:2007ec, Arnold:2008rz}.

A search for the SM Higgs boson is presented in the
$\PH \rightarrow \cPZ\cPZ \rightarrow 2\ell2\cPgn$ channel (where $\ell$ refers to either $\Pe$ or $\Pgm$), which is especially sensitive in
the high-mass range 250--600\GeV.  Results are reported from a
data sample corresponding to an integrated luminosity of 4.6\fbinv
recorded in 2011 by the Compact Muon Solenoid (CMS) experiment at $\sqrt{s}$ = 7\TeV.

\section{CMS Detector and Simulations}
A detailed description of the CMS detector can be found in
Ref.~\cite{CMSDETECTOR}. The key components of the detector include a
silicon pixel and a silicon strip tracker, embedded in a 3.8\unit{T}
solenoidal magnetic field, used to measure the momentum of
charged particles. The silicon pixel
and strip tracking system covers the pseudorapidity
range $|\eta| < 2.5$,  where $\eta=-\ln{[\tan{(\theta/2)}]}$, and $\theta$ is the polar angle of
the trajectory of the particle with respect to the beam direction.
It is surrounded by a crystal electromagnetic
calorimeter (ECAL) and a brass-scintillator hadron calorimeter
(HCAL). The ECAL and HCAL extend to a pseudorapidity range of $|\eta|
< 3.0$.  A steel/quartz-fiber Cherenkov forward detector (HF) extends
the calorimetric coverage to $|\eta| < 5.2$.
The calorimeters are surrounded by the muon system, used to identify
muons and measure their momentum. The muon system consists of gas detectors
placed in the steel return yoke of the magnet.

The largest background to the SM Higgs boson signal consists of events
in which a $\cPZ$ boson is produced in association with jets
($\cPZ+\text{jets}$). The $\cPZ+\text{jets}$ cross section is five orders of magnitude
larger than the expected production cross section for the signal.  The
other major backgrounds are top-quark production ($\ttbar\rightarrow
2\ell 2\cPgn2\cPqb$ and $\cPqt\PW\rightarrow 2\ell2\cPgn\cPqb$, and the
diboson production ($\PW\cPZ\rightarrow 3\ell\cPgn$, $\cPZ\cPZ\rightarrow 2\ell 2\cPgn$,
and $\PW\PW\rightarrow 2\ell 2\cPgn$).

Several Monte Carlo event generators are used to simulate the signal
and background processes.  The $\PH \rightarrow \cPZ\cPZ \rightarrow 2\ell
2\cPgn$ signal and top-quark background events are generated by using the
next-to-leading order (NLO) program \POWHEG2.0~\cite{POWHEG}.
The $\cPZ+\text{jets}$ and diboson backgrounds are simulated by using the \MADGRAPH5.1.3
generator~\cite{madgraph}.  The diboson backgrounds are also simulated using
the \PYTHIA6.4.22 generator for evaluating certain systematic uncertainties.
For events generated using \POWHEG and \MADGRAPH generators, parton showering
is simulated by using \PYTHIA with the Z2 tune which differs from the Z1 tune 
described in Ref.~\cite{Z1Ref} as it uses the CTEQ6~\cite{CTEQ6l} parametrization for the parton 
distribution functions instead of the CTEQ5~\cite{CTEQ5} parametrization.
The signal events are reweighted so that the
transverse momentum ($\PT$) distribution of the Higgs boson agrees with the
next-to-next-to-leading order (NNLO) and next-to-next-to-leading log
(NNLL) prediction~\cite{Bozzi:2005wk,deFlorian:2011xf}. The total
cross section is taken from Ref.~\cite{LHCHiggsCrossSectionWorkingGroup:2011ti}
and is scaled by the $\PH\rightarrow \cPZ\cPZ\rightarrow 2\ell2\cPgn$ branching ratio
~\cite{LHCHiggsCrossSectionWorkingGroup2,Djouadi:1997yw,hdecay2,Bredenstein:2006rh,Bredenstein:2006ha,Actis:2008ts}.
The parton distribution functions (PDF) are modeled through the CTEQ6L~\cite{CTEQ6l} parametrization
at leading order and the CT10 parametrization~\cite{Lai:2010vv} at NLO.
The NLO contribution to the $\Pq\Pq\rightarrow \cPZ\cPZ$ process is taken into account by reweighting the transverse
momentum of the visible \cPZ\ boson to match the prediction from the \textsc{mcfm 6.0} program~\cite{MCFM}.
A correction of 12\% of the leading order $\Pq\Pq\rightarrow \cPZ\cPZ$
cross section is included to account for the $\Pg\Pg\rightarrow \cPZ\cPZ$
process~\cite{gluonQCD}.  The detector
response to the simulated events is modeled with
\GEANTfour~\cite{GEANT4,CMSDETECTOR} and reconstruction and analysis
are performed by using the same software used for data.

\section{Event Selection}
For Higgs boson masses considered in this analysis,
the $\cPZ$ bosons from $\PH\rightarrow \cPZ\cPZ$ decay are typically produced
with a substantial $\PT$.
Events are therefore selected to have two well-identified, isolated,
opposite-charge leptons of same flavour ($\Pep\Pem$ or
$\Pgmp\Pgmm$) with $\PT > 20\GeV$ that have an
invariant mass within 30\GeV window centred on the $\cPZ$
mass.  The $\PT$ of the dilepton system is required to be greater than
55\GeV.  In the electron channel, these events are collected by using
dielectron triggers, with thresholds of $\PT > 17\GeV$ and $\PT > 8\GeV$
for the leading and the other electron, respectively.  The
muon channel relies on a combination of single- and double-muon triggers.
As instantaneous luminosity increased,
the thresholds on the double-muon triggers changed from a
requirement of $\PT > 7\GeV$ for each of the two muons to $\PT > 17\GeV$
and $\PT > 8\GeV$ on the leading and the other muon,
respectively.  The threshold for the single-muon trigger increased
from $\PT > 17\GeV$ to $\PT > 24\GeV$.  The trigger efficiency for
signal, for events selected through the full set of offline requirements
is measured by using $\cPZ$ decays in data, and ranges from 95\%
to 97\% in the muon channel and exceeds 99\% in the electron channel.

Muon candidates are reconstructed by using two algorithms, one in
which tracks in the silicon tracker are matched to energy deposits in
the muon detectors and another in which a combined fit is performed to
signals in both the silicon tracker and the muon
system~\cite{CMS-PAS-MUO-10-002}.  The muon candidates for
analysis are required to be successfully reconstructed through both
algorithms. Other identification criteria based on the number of
measurements in the tracker and in the muon system, the fit quality of
the muon track, and its consistency with the origin from the primary vertex are
also imposed on the muon candidates to reduce the misidentification rate.

Electron reconstruction also involves two
algorithms~\cite{CMS-PAS-EGM-10-004}, one in which energy clusters
in the ECAL are matched to signals in the silicon tracker and another
in which tracks in the silicon tracker are matched to ECAL
clusters. The electron candidates used in the analysis can be
reconstructed by either algorithm. More identification criteria
based on the distribution of the shower in the ECAL, a matching of the trajectory
of an electron track with the cluster in the ECAL, and
consistency with origin of the track from the primary vertex are imposed on the electron
candidates to reduce the misidentification rate. Electron candidates
with an ECAL cluster in the transition region between ECAL barrel and
endcap ($1.4442 < |\eta| < 1.566$) are rejected.  Additional
requirements are imposed to remove electrons produced in photon
conversions in the detector material.

Leptons produced in the decay
of $\cPZ$ bosons are expected to be isolated from hadronic activity in
the event. The sum of scalar transverse momentum depositions in the calorimeters and the transverse
momenta of tracks in a cone of radius $0.3$ in $\eta$--$\phi$ space around
each lepton, where $\phi$ is the azimuthal angle, is corrected by the contribution from the lepton
and the ratio of this corrected sum divided by the lepton $\PT$ is required to be
smaller than 15\%~(10\%) for muons (electrons).
To correct for the contribution to the isolation sum from pile-up interactions
(overlapping minimum-bias events from other concurrent proton-proton collisions),
a median energy density ($\rho$) is determined event by event~\cite{FASTJET}. Then the pile-up
contribution to the isolation sum is estimated as the product of $\rho$ and the area
of the cone in which the isolation sum is computed, and it is subtracted from the isolation sum
to make it largely insensitive to pile-up.
The combined reconstruction, identification and isolation efficiency
is measured in data by using $\cPZ$ decays and ranges between 90\% and 97\%
for muons, and between 70\% and 90\% for electrons, depending on the
$\PT$ and $\eta$ of the leptons.

The high instantaneous luminosity delivered by the LHC
provides an average of about 10 pile-up interactions
per bunch crossing, leading to events with several possible primary vertices.
The vertex with largest value of $\sum \PT^2$ for the associated
tracks is chosen to be the reference vertex.
According to simulation, this requirement provides the correct assignment for the primary
vertex in more than 99\% of both signal and background events.

The presence of a large imbalance in transverse momentum in an event
(\MET) is a fundamental feature of the signal.
The value of \MET is the modulus of the $\vec{E}_\mathrm{T}^\text{miss}$ vector computed as
the negative of the vector sum of the transverse momenta
of all reconstructed objects identified through the particle-flow algorithm,
which aims to reconstruct all particles produced in a collision event
by combining information from all sub-detectors~\cite{CMS-PAS-PFT-09-001}.
A large \MET threshold is imposed to suppress the bulk of the $\cPZ$+jets background,
which contains little genuine \MET.
The region of large \MET is populated by $\cPZ$+jets events in which the \MET
is largely due to jet mismeasurement.
To suppress the background with \MET arising from
mismeasurement of jets, events are removed if the angle in the azimuthal
plane between the \MET and the closest jet with transverse energy $\ET > 30\GeV$ is
smaller than 0.5 radians.  For events having no jets with $\ET > 30\GeV$,
this requirement is imposed between \MET and the closest jet with $\ET > 15\GeV$. Jets are reconstructed
from particle-flow candidates~\cite{CMS-PAS-PFT-09-001, CMS-PAS-JME-10-003} by using the anti-$\mathrm{k_{T}}$
clustering algorithm \cite{Cacciari:2008gp} with a distance parameter R of $0.5$, as implemented
in the {\sc fastjet} package~\cite{fastjetmanual, Cacciari:2006gp}.

Top-quark decays are characterized by the presence of jets originating from \cPqb\ quarks (\cPqb jets),
which are tagged on the basis of impact parameters of tracks in a jet, relative to
the primary vertex~\cite{CMS-PAS-BTV-09-001, CMS-PAS-BTV-11-001}.
The top-quark background is suppressed by applying a veto on events having a \cPqb\ tagged jet with
transverse energy greater than 30\GeV that lies within the tracker volume ($|\eta| < 2.4$).
To further suppress the top-quark background, a veto is applied on events containing a ``soft muon''
with $\PT > 3\GeV$, which is typically produced in the leptonic decay of a b quark.
The soft-muon veto along with the \cPqb-jet veto reduces the top-quark background by a factor of six.
To reduce the $\PW\cPZ$ background in which both bosons decay leptonically,
any event with a third lepton ($\Pe$ or $\Pgm$) with $\PT > 10\GeV$ and
passing the identification and isolation requirements is rejected.

\section{Analysis Strategy}
The search for the SM Higgs boson is performed
by using a transverse mass ($\MT$) variable as the final
discriminant in searching for an excess of events from the presence of the signal.
The transverse mass is defined as follows:

$$\MT^{2} = \left(\sqrt{{\PT(\ell\ell)}^{2} + {M(\ell\ell)}^{2}} + \sqrt{{\MET}^{2} +
{M(\ell\ell)}^{2}}\right)^{2} - (\vec{p}_{T}(\ell\ell) + \vec{E}_T^\text{miss})^{2}.$$

where $\PT(\ell\ell)$ and $M(\ell\ell)$
are the transverse momentum and invariant mass of dilepton system, respectively.

Two approaches are considered:
a ``cut-based'' and a ``shape-based'' analysis.
The same set of event selection criteria is used by both the
analyses, except for requirements on the \MET and $\MT$ variables.
In both cases, requirements on \MET and $\MT$ vary with the Higgs mass hypothesis, but
are typically more relaxed for the shape-based analysis.

In the cut-based analysis, the Higgs boson search is performed by
looking for an excess of events above the standard model background
expectation after applying selection criteria that are tuned for a
given Higgs boson mass hypothesis.  The \MET and $\MT$ selection
listed in Table~\ref{tab:met_mtcuts} was optimized by using a genetic
algorithm for rectangular cuts optimization, \textsc{
garcon}~\cite{GARCON}, with the expected 95\% CL limit
on the cross section normalized to the standard model Higgs boson cross section as a figure of merit.
Optimization was performed on six $\mH$ points, and a smooth-fit interpolation for the cut
values was performed to provide optimal cut values for every explored
Higgs mass point.

In the shape-based analysis, the Higgs boson search is performed by using
a binned likelihood fit to the $\MT$ distribution obtained after
applying the selection given in Table~\ref{tab:met_mtcuts}. The range of $\MT$ is varied
as a function of $\mH$ to ensure that the expected signal distribution is fully contained.
The shape-based analysis is found to have the best expected signal
sensitivity and is considered as the main result in the paper.

\begin{table}[htp]
\caption{
      Higgs boson mass-dependent selection for \MET and
      $\MT$ variables in the cut-based and shape-based analyses.}
\begin{center}
\begin{tabular}{c|cccccc}\hline
$\mH$ (\GeVns)    & 250        & 300        & 350        & 400       & 500       & 600 \\
\hline
\multicolumn{7}{c} {Cut-based analysis selection} \\
\hline
\MET (\GeVns)     &      $> 70$   &     $> 79$   &    $> 95$    & $> 115$    & $> 150$       & $> 161$ \\
$M_T$ (\GeVns)   & [222, 272] & [264, 331] & [298, 393] & [327, 460] & [382, 605] & [452, 767] \\
\hline
\multicolumn{7}{c} {Shape-based analysis selection} \\
\hline
\MET (\GeVns)     & $> 70$         &  $> 80$        & $> 80$         &  $> 80$    & $> 80$         & $> 80$ \\
$M_T$ (\GeVns)   & [180, 300] & [250, 350] & [250, 400] & [250, 450]  & [250, 600] &[250, 750] \\
\hline
\end{tabular}
\end{center}
\label{tab:met_mtcuts}
\end{table}

\section{Background Estimation}
\label{sec:bkgest}

For both types of analyses,
the $\cPZ\cPZ$ and $\PW\cPZ$ backgrounds are modeled using Monte Carlo simulation, and are
normalized to their respective NLO cross sections. The remaining backgrounds are
estimated using control samples in data.

\begin{figure}[ht]
{\begin{center}
\includegraphics[width=0.48\textwidth]{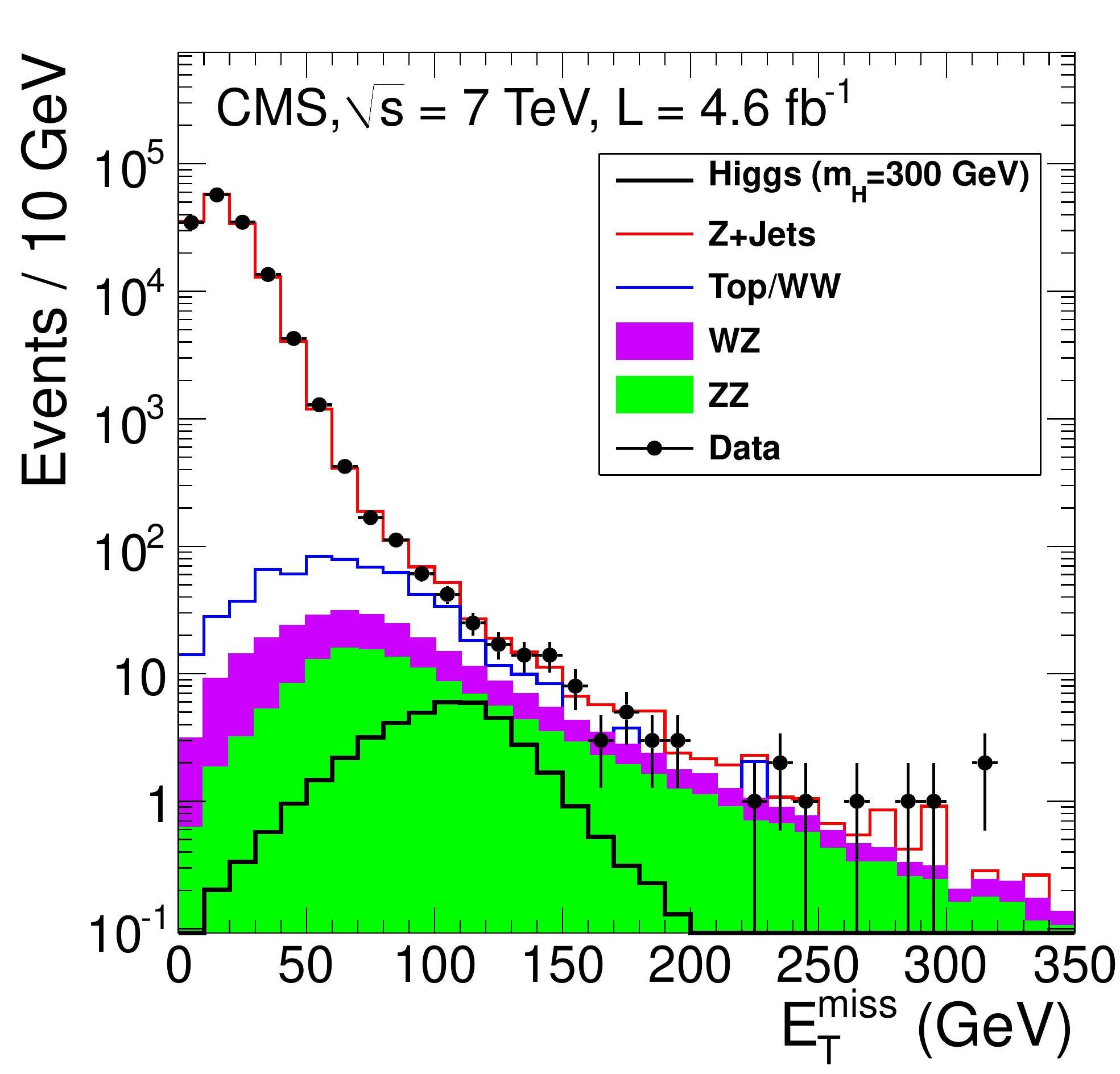}
\end{center}
\caption{The \MET distribution in data compared to the background simulation.
The dielectron and dimuon channels are combined.
Contributions from $\cPZ\cPZ$, $WZ$, non-resonant background and $\cPZ$+jets background are stacked
on top of each other. The \MET distribution in signal events for $\mH = 300$\GeV is also shown.}
\label{fig:zgamma_met_data}
}
\end{figure}

The $\cPZ$+jets background is simulated from a control sample of events
with a single photon produced in association with jets ($\Pgg+\text{jets}$).
This obviates the need for using less reliable Monte Carlo simulation of
the \MET distribution arising from mismeasurement of jets.
The $\Pgg+\text{jets}$ cross section is much larger than $\cPZ+\text{jets}$.
The jets must have a minimum $\ET$ of 15\GeV
to reduce contamination from processes that have a photon
produced in association with genuine \MET,
such as $\PW(\ell\cPgn)+\Pgg$, $\PW(\ell\cPgn)$+jets, where the jet is mismeasured as a photon,
and $\cPZ(\cPgn\cPgn)+\Pgg$ events.
The kinematics and overall normalization of $\Pgg+\text{jets}$ events
are matched to $\cPZ+\text{jets}$ in data through an event-by-event
reweighting as a function of the boson $\PT$, and the number of jets with $\ET>30$\GeV,
to account for the dependence of the \MET on the associated hadronic activity.
Residual differences from pile-up are taken into account by reweighting events according
to the number of reconstructed vertices.
This procedure yields an accurate model of the \MET distribution in $\cPZ$+jets events,
as shown in Fig.~\ref{fig:zgamma_met_data}, which compares the
\MET distribution of the reweighted $\Pgg$+jets events along with other backgrounds to the \MET
distribution of the dilepton events in data.
To compute the $\MT$ for each
$\Pgg$+jets event, the value of $\vec{p}_\mathrm{T}(\ell\ell)$ is defined as the photon $\vec{p}_\mathrm{T}$
and the value of $M(\ell\ell)$ term is chosen according to a probability density function
constructed from the measured dilepton invariant mass distribution in $\cPZ$+jets events.

The background processes that do not involve a $\cPZ$ resonance (non-resonant background)
are estimated by using a control sample of events with dileptons of different flavor
($\Pe^{\pm}\Pgm^{\mp}$) that pass the full analysis selection.
This background consists mainly of leptonic \PW\ decays in $\ttbar$, $\cPqt\PW$ decays and
$\PW\PW$ events. Small contributions from single top-quark events produced from
$s$-channel and $t$-channel processes, $\PW$+jets events in which the $\PW$ boson
decays leptonically and a jet is mismeasured as a lepton, and $\cPZ\rightarrow \Pgt\Pgt$
events in which $\Pgt$ leptons produce light leptons and \MET are included in this estimate of the non-resonant background.
This method cannot distinguish between the non-resonant background and
a possible contribution from $\PH \rightarrow \PW\PW \rightarrow 2\ell 2\cPgn$ events, which are treated
as part of the non-resonant background estimate.

The non-resonant background in the $\Pep\Pem$ and $\Pgmp\Pgmm$ final states
is estimated by applying a scale factor ($\alpha$) to the selected
$\Pe^{\pm}\Pgm^{\mp}$ events:

\begin{equation*}
N_{\Pgm\Pgm} = \alpha_{\Pgm} \times N_{\Pe\Pgm}, \qquad
N_{\Pe\Pe} = \alpha_{\Pe} \times N_{\Pe\Pgm}.
\end{equation*}

This $\alpha$ factor is computed from the sidebands (SB) of the $\cPZ$ peak
(40 $< M(\ell\ell) <$ 70\GeV and 110~ $< M(\ell\ell) <$200\GeV) by using the following relations:

\begin{equation*}
\alpha_{\Pgm} = \frac{N_{\Pgm\Pgm}^\mathrm{SB}}{N_{\Pe\Pgm}^\mathrm{SB}}, \qquad
\alpha_{\Pe} = \frac{N_{\Pe\Pe}^\mathrm{SB}}{N_{\Pe\Pgm}^{SB}},
\end{equation*}

where $N_{\Pe\Pe}^\mathrm{SB}$, $N_{\Pgm\Pgm}^\mathrm{SB}$, and $N_{\Pe\Pgm}^\mathrm{SB}$ are the number of events
in the $\cPZ$ sidebands in a top-enriched sample of $\Pep\Pem$, $\Pgmp\Pgmm$, and $\Pe^{\pm}\Pgm^{\mp}$
final states, respectively. Such samples are selected by requiring \MET$>$70\GeV and a \cPqb-tagged jet in the events.
The measured values of $\alpha$ with the corresponding statistical uncertainties
are $\alpha_{\Pgm}= 0.58 \pm 0.02$ and $\alpha_{\Pe} = 0.42 \pm 0.02$.

Since the $\MT$ distribution obtained for the
non-resonant background is based on a small sample of $\Pe^{\pm}\Pgm^{\mp}$ events scaled by $\alpha$,
this distribution is smoothed~\cite{Smoothing} for use in the shape-based analysis.

\section{Systematic Uncertainties}
Systematic uncertainties include experimental uncertainties
on the selection and measurement of the reconstructed objects,
theoretical uncertainties on the signal and background
processes which are derived from Monte Carlo simulation, and uncertainties on
backgrounds determined from control samples in data.
These are summarised in Table~\ref{tab:syst_summary}.

The theoretical uncertainties on the predicted signal are evaluated
by changing the parton distribution functions and the
QCD renormalization and factorization scales~\cite{LHCHiggsCrossSectionWorkingGroup:2011ti,Botje:2011sn,Alekhin:2011sk,Lai:2010vv,Martin:2009iq,Ball:2011mu}.
The theoretical uncertainties on the NLO cross sections of the
$\cPZ\cPZ$ and $\PW\cPZ$ backgrounds are estimated by the same procedure of
changing the parton distribution functions and the QCD renormalization and factorization scales
by using the program \textsc{mcfm} 6.0.

The cross section for on-shell Higgs boson production and decay is calculated in the zero-width
approximation. Recent studies show that current Monte Carlo simulation does not describe
the correct Higgs-boson line shape for $ \mH \gtrsim 300$\GeV. These effects are estimated
to correspond to an additional uncertainty of 10--30\% on the theoretical cross section for
$400 < \mH < 600$\GeV~\cite{Passarino:2010qk,LHCHiggsCrossSectionWorkingGroup:2011ti,Anastasiou:2011pi}.

Various factors contribute to the experimental uncertainties that apply to
processes derived from Monte Carlo simulation.
These include uncertainties on the trigger efficiency and lepton selection efficiencies.
The effect of lepton momentum scale and jet energy scale is also taken
into account and is propagated to the evaluation of \MET.
The uncertainty on the b jet veto is estimated by measuring the b-tagging
efficiency in data from fully leptonic $\ttbar$ decays~\cite{CMS-PAS-BTV-11-001}.
The uncertainty due to the modeling of pile-up is evaluated by shiftng the mean of the distribution
of the number of pile-up interactions up and down by one.

The uncertainty on the estimate of the
non-resonant background is dominated by the statistical uncertainty
of the $\Pe^{\pm}\Pgm^{\mp}$ control sample.
The uncertainty on the prediction of the $\cPZ$+jets background is affected by any
residual contamination of the $\Pgg$+jets control sample from processes involving
a photon and genuine $\MET$. Consequently, this method provides only an upper
bound on the prediction. To account for this contamination, the $\cPZ$+jets
background prediction is changed from zero to the estimate obtained
from the $\Pgg$+jets control sample in the limit-setting procedure.
This is achieved by taking half of the estimate from the $\Pgg$+jets control sample
as the central value for the $\cPZ$+jets background and assigning a 100\% systematic uncertainty
to this value. Since $\cPZ$+jets contributes to only 10--15\% of the total background,
the large uncertainty on this background does not have a significant impact on the overall
performance of the analysis.

In the shape-based analysis,
the difference in the $\MT$ distribution with respect to an alternate Monte Carlo
generator (\textsc{madgraph} versus \PYTHIA) is used as an uncertainty in the modeling of the $\cPZ\cPZ, \PW\cPZ$ backgrounds.
The uncertainty on the $\MT$ distributions arising from
the limited number of Monte Carlo simulated events is taken into account by
scaling all the bins up and down by the respective statistical uncertainties.
This is done in a correlated manner to obtain two bounding
distributions, within which the nominal distribution is changed.
For the non-resonant background, where the nominal
$\MT$ distribution is obtained by smoothing the $\MT$ distribution of $\Pe^{\pm}\Pgm^{\mp}$ events,
the bounding distributions are taken from simulation to account for any possible bias
introduced by smoothing.
The uncertainty for signal is assigned by
reweighting the $\PT$ spectrum of the Higgs boson to the one
obtained from changing the renormalization
and factorization scales in the NNLO+NNLL calculation~\cite{Bozzi:2005wk,deFlorian:2011xf}.

\begin{table}[htb]
  \caption{Summary of systematic uncertainties on event yields of signal and background processes.
Ranges refer to uncertainties that depend on the hypothesised value of $\mH$.}
  \begin{center}
    \begin{tabular}{|l|l|}
        \hline
        Source                                       & Uncertainty $[\%]$           \\ \hline\hline
        Luminosity                                   & 4.5                          \\ \hline
        PDF for Higgs production (gluon-fusion)      & 8.2--10.4                       \\ \hline
        PDF for Higgs production (VBF)               & 3.8--7.6                       \\ \hline
        PDF for $\Pq\Pq\rightarrow \cPZ\cPZ$                   & 4.8                           \\ \hline
        PDF for $\Pq\Pq\rightarrow \PW\cPZ$                   & 5.5                           \\ \hline
        QCD scale for Higgs production (gluon-fusion)& 7.6--11.1                        \\ \hline
        QCD scale for Higgs production (VBF)         & 0.4--2                        \\ \hline
        QCD scale for $\Pg\Pg\rightarrow \cPZ\cPZ$             & 20                           \\ \hline
        QCD scale for $\Pq\Pq\rightarrow \cPZ\cPZ$             & 6.2                           \\ \hline
        QCD scale for $\Pq\Pq\rightarrow \PW\cPZ$             & 8.5                           \\ \hline
        Higgs boson line shape                       & 10--30                        \\ \hline
        Trigger                                      & 1 (for \Pe\Pe), 2 (for $\Pgm\Pgm$) \\ \hline
        Lepton identification and isolation          & 2                            \\ \hline
        Lepton momentum scale                        & 5 (for \Pe\Pe), 2 (for $\Pgm\Pgm$) \\ \hline
        Jet energy scale                             & 1--1.5                        \\ \hline
        \cPqb jet Veto                                   & 1--1.2                        \\ \hline
        Pile-up                                      & 1--3                          \\ \hline
        Non-resonant background                      & 15--100                       \\ \hline
        \cPZ+jets                                       & 100                          \\ \hline
    \end{tabular}
  \end{center}
  \label{tab:syst_summary}
\end{table}

\section{Results}

\begin{figure}[ht]
{\centering
\includegraphics[width=0.48\textwidth]{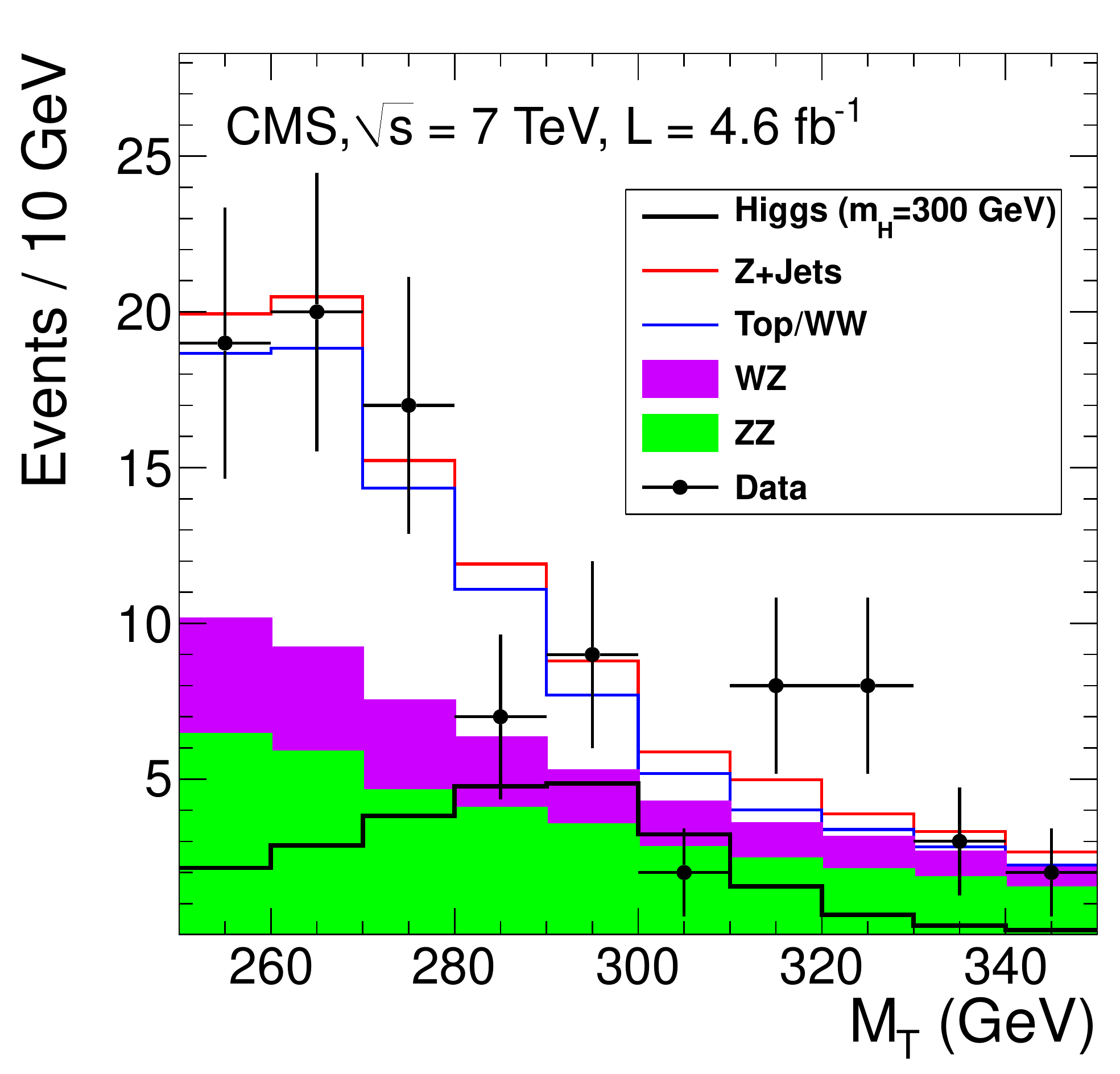}
\includegraphics[width=0.48\textwidth]{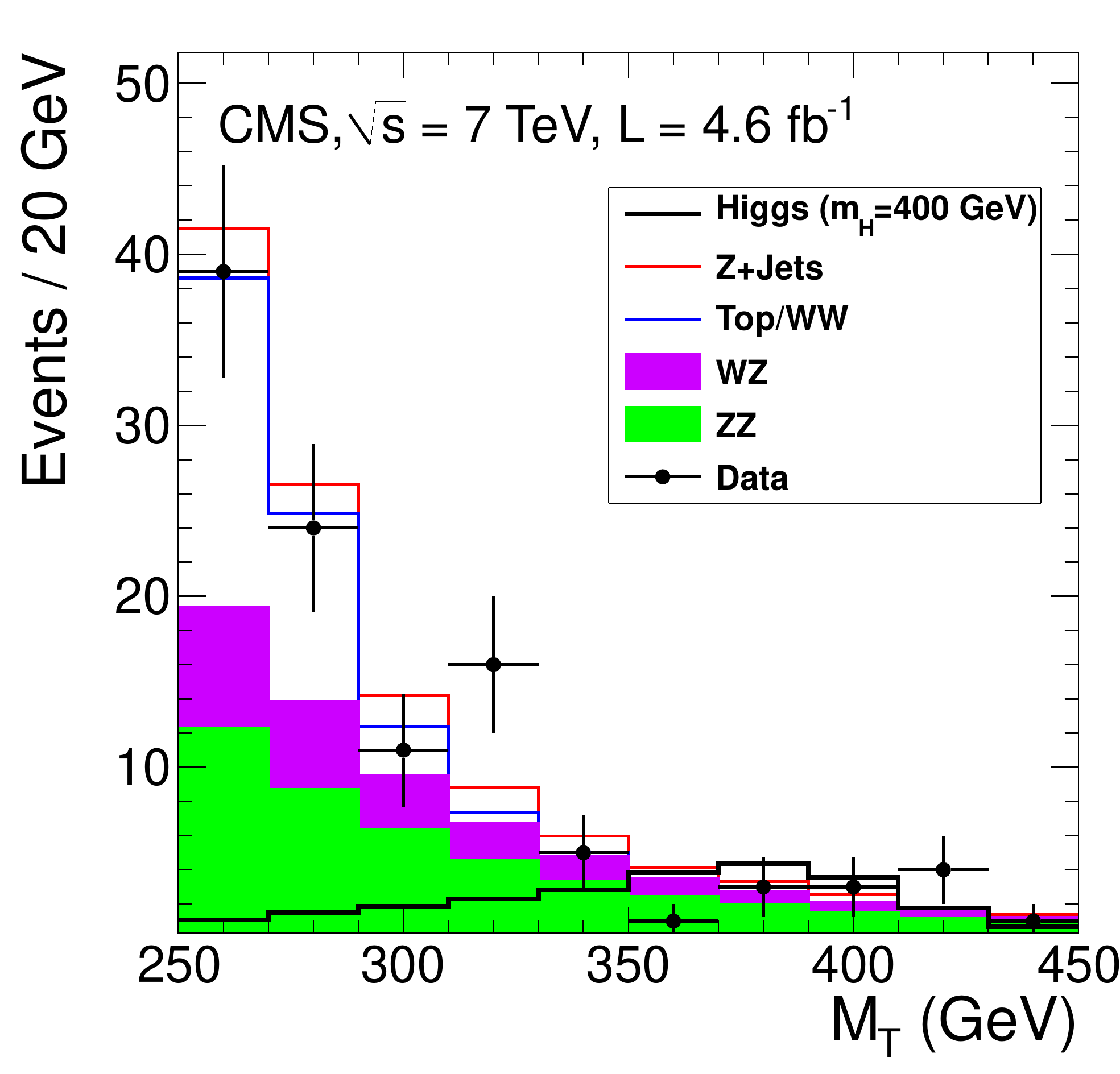}
\caption{The $\MT$ distribution for events passing $\mH$ selections for 300\GeV (left) and 400\GeV (right).
The dielectron and dimuon channels are combined.}
\label{fig:mt_fullsel}
}
\end{figure}

The event yields for the cut-based analysis, as a function of $\mH$,
and the estimated backgrounds for
the 4.6\fbinv dataset are listed in Table~\ref{tab:final_yields}.
The $\MT$ distributions for the shape-based analysis based on the selections
described in Table~\ref{tab:met_mtcuts} for $\mH = 300$ and 400\GeV hypotheses
are shown in Fig.~\ref{fig:mt_fullsel}.
No significant excess of events is observed over the expectation from the SM background
and limits are set on the production cross section
of the standard model Higgs boson as a function of $\mH$.

\begin{table}[ht]
\caption{Background estimates,
signal predictions, and observed number of events for an integrated luminosity of
4.6\fbinv after the cut-based selection. The uncertainties
include the statistical and systematic components (in that order).
The results for signal contain only the systematic uncertainties as the statistical uncertainties are negligible.
For the non-resonant background, the 68\% upper limit is quoted in the case where the background prediction is zero.}

\begin{center}
\scriptsize\begin{tabular}{|l|c|c|c|c|c|c|c|} \hline
$\mH$                 & \cPZ\cPZ            & WZ             & Top/\PW\PW/               & \cPZ+Jets         & Total     & Expected         & Data   \\
(\GeVns)           &               &                & \PW+jets/\cPZ$\rightarrow \Pgt\Pgt$ &       &   Background     & Signal            &        \\ \hline
\hline

$250$ & $36.0\pm 0.2\pm 2.6$ & $24.0\pm 0.3\pm 2.0$ & $65.0\pm 3.8\pm 5.8$ & $15.0\pm 15.0$ & $140.0\pm 3.8\pm 16.0$ & $22.0\pm 2.2$ & $142$ \\ \hline
$300$ & $23.0\pm 0.2\pm 1.7$ & $13.0\pm 0.2\pm 1.1$ & $18.0\pm 1.1\pm 3.0$ & $6.3\pm 6.3$ & $60.0\pm 1.1\pm 7.3$ & $21.0\pm 2.1$ & $64$ \\ \hline
$350$ & $16.0\pm 0.1\pm 1.1$ & $7.0 \pm 0.2\pm 0.6$ & $2.0 \pm 0.1\pm 1.0$ & $4.1\pm 4.1$ & $29.0\pm 0.3\pm 4.4$ & $21.0\pm 2.5$ & $26$ \\ \hline
$400$ & $12.0\pm 0.1\pm 0.9$ & $4.6 \pm 0.1\pm 0.4$ & $< 1.1$              & $2.7\pm 2.7$ & $19.0\pm 0.2\pm 2.9$ & $17.0\pm 2.0$ & $18$ \\ \hline
$500$ & $7.5\pm 0.1\pm 0.5$  & $2.0 \pm 0.1\pm 0.2$ & $< 1.1$              & $1.4\pm 1.4$ & $11.0\pm 0.1\pm 1.5$ & $7.4\pm 1.3$ & $14$ \\ \hline
$600$ & $3.9\pm 0.1\pm 0.3$  & $0.8 \pm 0.1\pm 0.1$ & $< 1.1$              & $0.6\pm 0.6$ & $5.3\pm 0.1\pm 0.7$  & $2.9\pm 0.7$ & $5$ \\ \hline

\end{tabular}
\end{center}
\label{tab:final_yields}
\end{table}

\begin{figure}[ht]
{\centering
\includegraphics[width=0.49\textwidth]{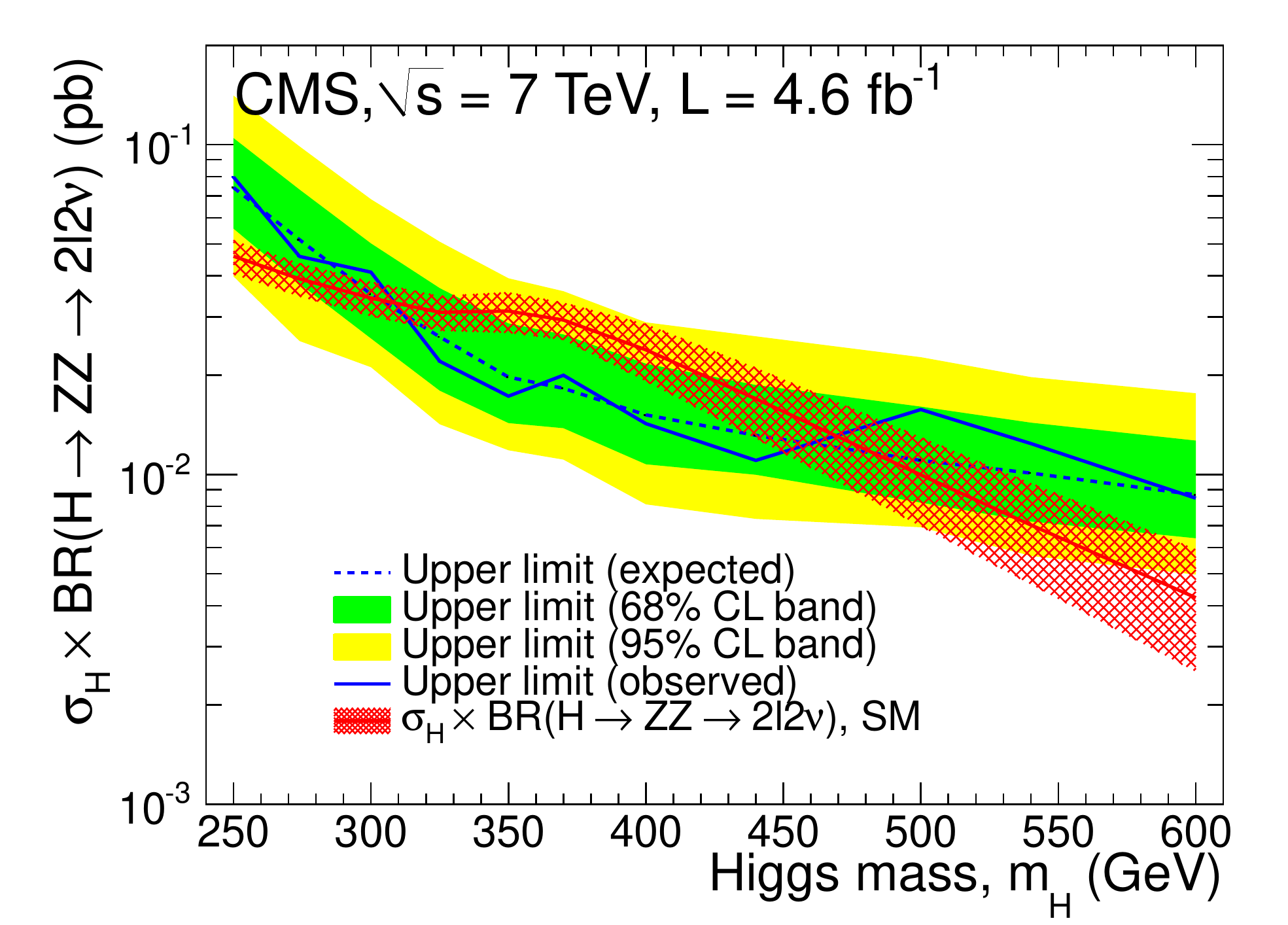}
\includegraphics[width=0.49\textwidth]{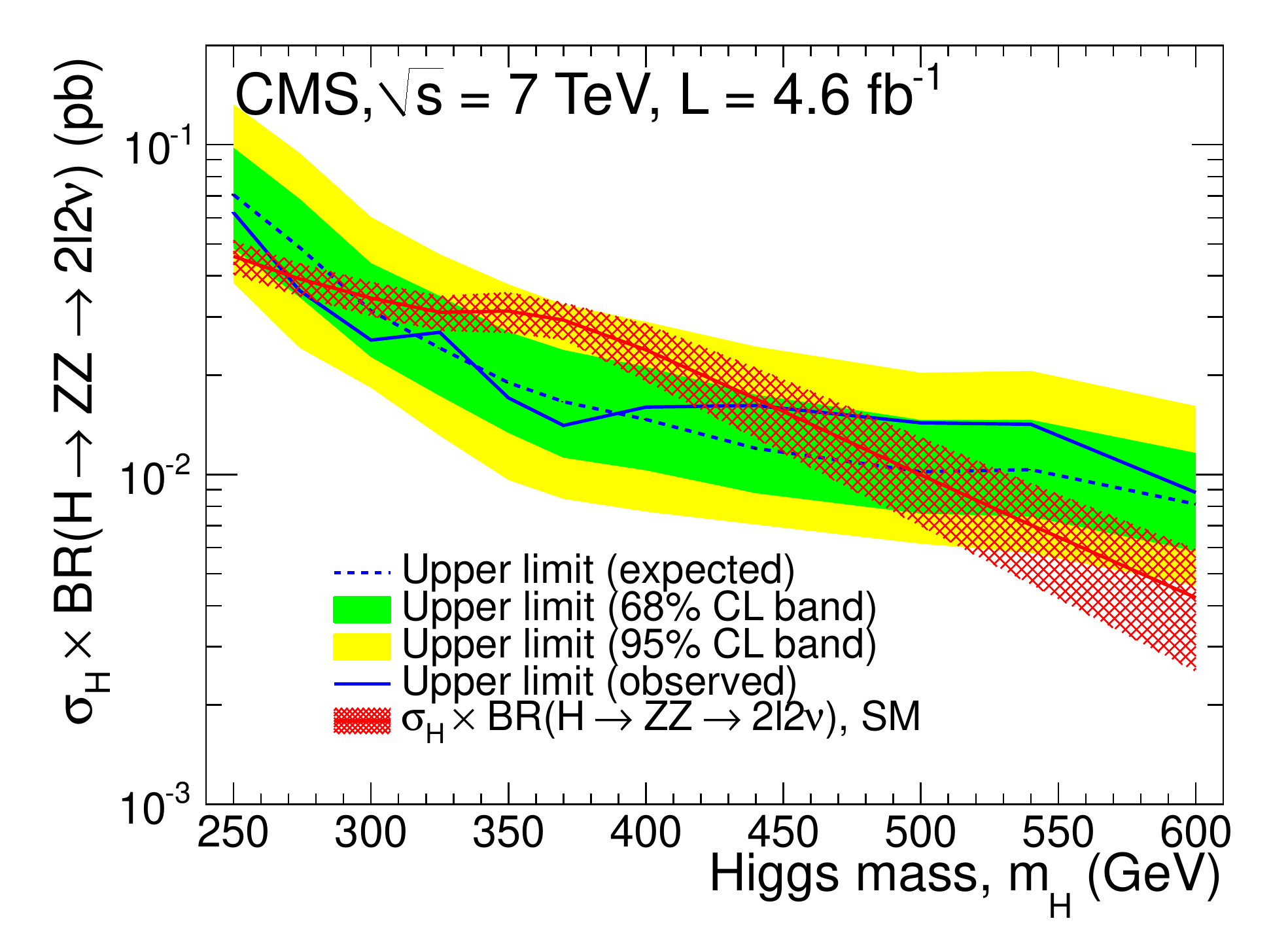}
\caption{
The median expected and observed 95\% CL upper limits on the cross section
$\sigma \times BR(H\rightarrow \cPZ\cPZ \rightarrow 2\ell2\cPgn)$ for the Higgs boson masses in
the range $250 - 600$\GeV for the cut-based (left) and shape-based
(right) analyses.}
\label{fig:xs_SM}
}
\end{figure}

\begin{figure}[ht]
{\centering
\includegraphics[width=0.49\textwidth]{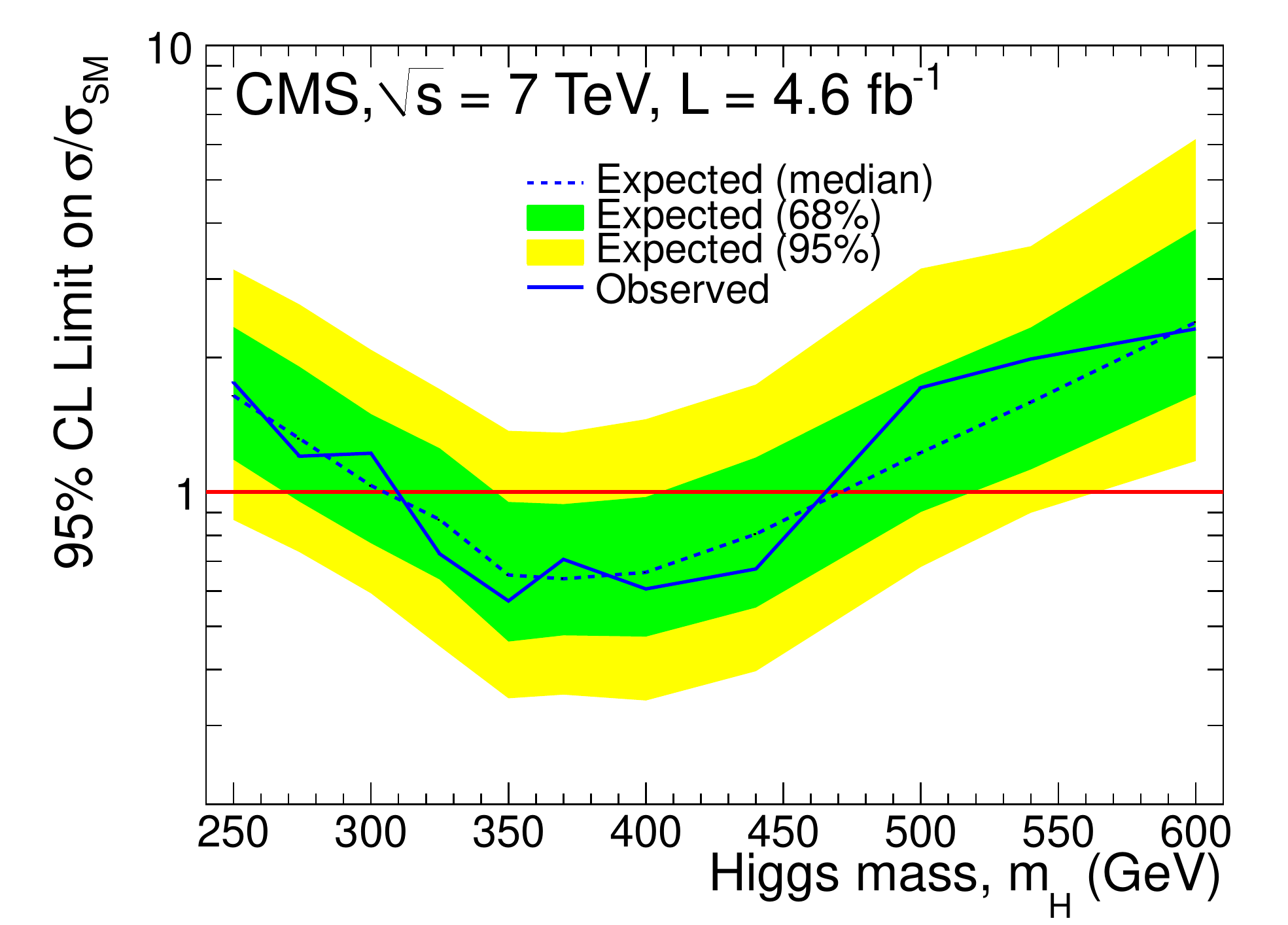}
\includegraphics[width=0.49\textwidth]{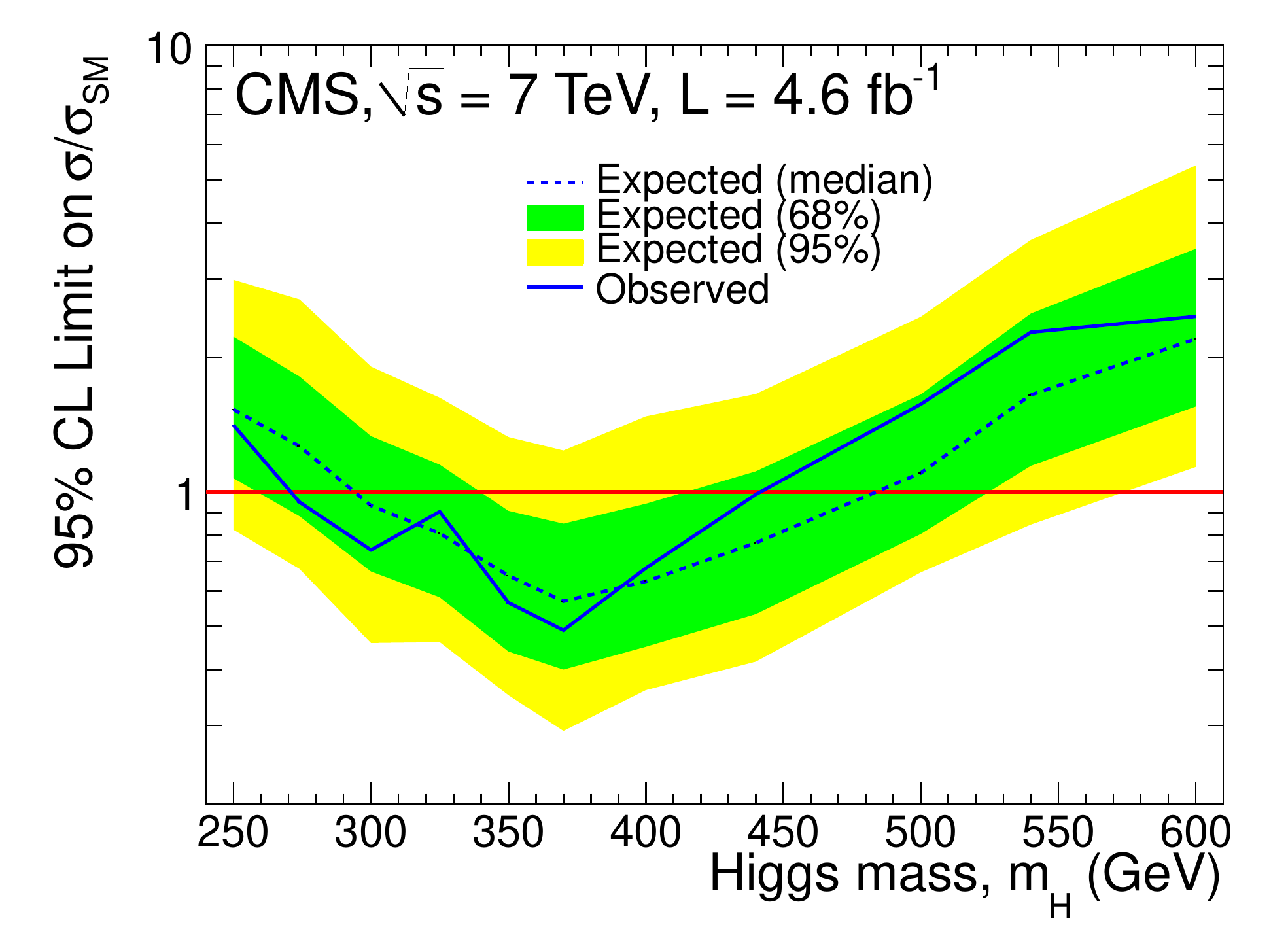}
\caption{
The ratio $R$ of the 95\% CL cross section upper limit $\sigma$ to the
SM Higgs boson production cross section $\sigma_\mathrm{SM}$ as a function of the
Higgs boson mass $\mH$ for the
cut-based (left) and shape-based (right) analyses.}
\label{fig:limits_SM}
}
\end{figure}

The median expected and observed
95\% CL upper limits on the cross section $\sigma \times BR(\PH\rightarrow
\cPZ\cPZ \rightarrow 2\ell 2\cPgn)$ for $250 < \mH < 600\GeV$
by using the cut- and the shape-based approaches are shown in Fig.~\ref{fig:xs_SM}.
The measured ratio $R$ of the 95\%
CL upper limit cross section $\sigma$ to the SM Higgs boson production cross section
$\sigma_\mathrm{SM}$ as a function of $\mH$ is shown in Fig.~\ref{fig:limits_SM}.
These results are obtained by using the CL$_\mathrm{s}$ method~\cite{Read2,TJunk,ATLASCMS}.
For the cut-based analysis the SM Higgs boson is excluded
in the mass range 310--465\GeV at 95\% confidence level, while the
expected exclusion limit in the background-only hypothesis is 305--470\GeV.
For the shape-based analysis the SM Higgs boson is excluded
in the mass range 270--440\GeV at 95\% confidence level while the
expected exclusion limit for the background-only hypothesis is 290--490\GeV.

\section{Summary}
A search for the standard model Higgs boson has been performed in the decay channel $\PH \rightarrow \cPZ\cPZ \rightarrow 2\ell 2\nu$
in pp collisions at $\sqrt{s} = $ 7\TeV, using a data sample corresponding
to an integrated luminosity of 4.6\fbinv. No significant excess
is found above the background expectation.
The presence of the SM Higgs boson is excluded for $270\GeV < \mH < 440$\GeV at 95\% CL.

\section*{Acknowledgements}

We congratulate our colleagues in the CERN accelerator departments for the excellent performance of the LHC machine. We thank the technical and administrative staff at CERN and other CMS institutes, and acknowledge support from: FMSR (Austria); FNRS and FWO (Belgium); CNPq, CAPES, FAPERJ, and FAPESP (Brazil); MES (Bulgaria); CERN; CAS, MoST, and NSFC (China); COLCIENCIAS (Colombia); MSES (Croatia); RPF (Cyprus); MoER, SF0690030s09 and ERDF (Estonia); Academy of Finland, MEC, and HIP (Finland); CEA and CNRS/IN2P3 (France); BMBF, DFG, and HGF (Germany); GSRT (Greece); OTKA and NKTH (Hungary); DAE and DST (India); IPM (Iran); SFI (Ireland); INFN (Italy); NRF and WCU (Korea); LAS (Lithuania); CINVESTAV, CONACYT, SEP, and UASLP-FAI (Mexico); MSI (New Zealand); PAEC (Pakistan); MSHE and NSC (Poland); FCT (Portugal); JINR (Armenia, Belarus, Georgia, Ukraine, Uzbekistan); MON, RosAtom, RAS and RFBR (Russia); MSTD (Serbia); MICINN and CPAN (Spain); Swiss Funding Agencies (Switzerland); NSC (Taipei); TUBITAK and TAEK (Turkey); STFC (United Kingdom); DOE and NSF (USA). Individuals have received support from the Marie-Curie programme and the European Research Council (European Union); the Leventis Foundation; the A. P. Sloan Foundation; the Alexander von Humboldt Foundation; the Belgian Federal Science Policy Office; the Fonds pour la Formation \`a la Recherche dans l'Industrie et dans l'Agriculture (FRIA-Belgium); the Agentschap voor Innovatie door Wetenschap en Technologie (IWT-Belgium); the Council of Science and Industrial Research, India; and the HOMING PLUS programme of Foundation for Polish Science, cofinanced from European Union, Regional Development Fund.

\bibliography{auto_generated}   % will be created by the tdr script.

\cleardoublepage \appendix\section{The CMS Collaboration \label{app:collab}}\begin{sloppypar}\hyphenpenalty=5000\widowpenalty=500\clubpenalty=5000\textbf{Yerevan Physics Institute,  Yerevan,  Armenia}\\*[0pt]
S.~Chatrchyan, V.~Khachatryan, A.M.~Sirunyan, A.~Tumasyan
\vskip\cmsinstskip
\textbf{Institut f\"{u}r Hochenergiephysik der OeAW,  Wien,  Austria}\\*[0pt]
W.~Adam, T.~Bergauer, M.~Dragicevic, J.~Er\"{o}, C.~Fabjan, M.~Friedl, R.~Fr\"{u}hwirth, V.M.~Ghete, J.~Hammer\cmsAuthorMark{1}, M.~Hoch, N.~H\"{o}rmann, J.~Hrubec, M.~Jeitler, W.~Kiesenhofer, M.~Krammer, D.~Liko, I.~Mikulec, M.~Pernicka$^{\textrm{\dag}}$, B.~Rahbaran, C.~Rohringer, H.~Rohringer, R.~Sch\"{o}fbeck, J.~Strauss, A.~Taurok, F.~Teischinger, P.~Wagner, W.~Waltenberger, G.~Walzel, E.~Widl, C.-E.~Wulz
% ...
\vskip\cmsinstskip
\dag:~Deceased\\
1:~~Also at CERN, European Organization for Nuclear Research, Geneva, Switzerland\\
% ...

\end{sloppypar}
\end{document}